# Large voltage-induced modification of spin-orbit torques in Pt/Co/GdOx

Satoru Emori,[*a)] Uwe Bauer, Seonghoon Woo, Geoffrey S. D. Beach[b)]

Department of Materials Science and Engineering, Massachusetts Institute of Technology,

Cambridge, Massachusetts 02139, USA

We report on large modifications of current-induced spin-orbit torques in a gated Pt/Co/Gd-oxide microstrip due to voltage-driven $O^{2-}$ migration. The Slonczewski-like and field-like torques are quantified using a low-frequency harmonic technique based on the polar magneto-optical Kerr effect. Voltage-induced oxidation of Co enhances the Slonczewski-like torque by as much as an order of magnitude, and simultaneously reduces the anisotropy energy barrier by a factor of ~5. Such magneto-ionic tuning of interfacial spin-orbit effects may significantly enhance the efficiency of magnetization switching and provide additional degrees of freedom in spintronic devices.

*Present address:

Department of Electrical and Computer Engineering, Northeastern University, Boston, Massachusetts 02115, USA

e-mail: a) s.emori@neu.edu, b) gbeach@mit.edu

Electrical control of magnetization is essential for realizing high-performance, low-power spintronic memory and logic devices. In heavy-metal/ferromagnet/oxide trilayers with strong spin-orbit coupling, Rashba and spin Hall effects can give rise to current-induced torques[1] on the ferromagnetic film that can efficiently drive magnetization switching[2–7] and domain-wall motion.[8–12] These spin-orbit torques (SOTs)[13–15] exhibit two symmetries: a Slonczewski-like (or damping-like) torque that counteracts damping and directly drives magnetization switching,[2,16] and a field-like torque that can lower the energy barrier for switching.[17] It is anticipated that devices based on SOTs may outperform those based on conventional spin-transfer torque.[18]

Interfacial spin-orbit coupling can also generate perpendicular magnetic anisotropy (PMA) in ultrathin ferromagnets, which is required to successfully scale devices to smaller lateral dimensions. Strong PMA is necessary to maintain a sufficiently large energy barrier against thermal fluctuations. At the same time, since the anisotropy barrier determines the critical threshold for current-induced switching, it is desirable to lower PMA during switching and recover it afterwards. Voltage control of interfacial PMA has already been demonstrated based on charge accumulation at a ferromagnet/dielectric interface[19–22] and used for voltage-assisted spin-torque switching in magnetic tunnel junctions.[23–25] Since SOTs also derive from interfacial spin-orbit coupling, SOTs may likewise be amenable to gate voltage control. Indeed, a recent study has demonstrated that voltage-induced interfacial charging can measurably influence SOTs in Pt/Co/$Al_2O_3$.[26] However, since the charge density in a metal is difficult to change significantly, effects based on voltage-induced electron accumulation are typically very small.

We have recently reported that large nonvolatile modification of interfacial magnetism in ultrathin Co, sandwiched between Pt and Gd-oxide (GdOx), can be achieved through ion accumulation at the Co/GdOx interface.[27–29] This "magneto-ionic" effect, much stronger than conventional magneto-electric effects, relies on the motion of oxygen ions in GdOx and a consequent modification of the oxidation state at the Co/GdOx interface.[29] In this Letter, we demonstrate that a substantial voltage-induced modification of SOTs in Pt/Co/GdOx can be achieved through voltage-driven oxygen migration. Voltage-induced interfacial Co oxidation simultaneously decreases the anisotropy energy barrier significantly, so that magneto-ionic control of interfacial magnetism should dramatically enhance the SOT switching efficiency. More broadly, our results suggest that voltage control of interface chemistry and structure provides a powerful means to probe the fundamental origins of SOTs and other spin-orbit-derived interfacial phenomena.

Ta(4nm)/Pt(3nm)/Co(0.9nm)/GdOx(3nm) films were sputter-deposited onto Si/SiO$_2$(50 nm) substrates at room temperature, where the GdOx layer was grown reactively from a metallic Gd target under an oxygen partial pressure of ~5×10$^{-5}$ Torr. Gated 6 μm wide microstrip devices (Fig. 1(a)) were fabricated using electron-beam lithography and liftoff. After fabricating the strips, Ta/Cu electrodes were patterned at either end for current injection, and then a 10 μm wide GdOx(30nm)/Ta(2nm)/Au(12nm) gate was placed across the strip using two additional lithographic steps (Fig. 1(b)). The device in the virgin state exhibits strong PMA, as evidenced by a square out-of-plane hysteresis loop (Fig. 1(c)) and a large anisotropy field $H_k \approx 6000$ Oe, measured using the polar magneto-optical Kerr effect (MOKE), as described below.

Polar MOKE measurements were performed using a scanning Kerr polarimeter, with a ~3 μm laser spot positioned with a high-resolution scanning stage. The laser spot was smaller than the gate dimensions, and the Au electrode was sufficiently thin, so that the magnetic properties could be directly probed underneath the center of the gate region. $H_k$ was determined by measuring the polar MOKE signal (proportional to the z-component of magnetization, $M_z$) versus in-plane field $H_x$, which in the Stoner-Wohlfarth model follows $M_z/M_s = \cos[\arcsin(H_x/H_k)]$. To improve the signal to noise ratio, periodic out-of-plane field pulses of alternating polarity were applied to toggle the z-component of the canted magnetization between $\pm M_z$. The MOKE signal was then measured with a lock-in amplifier phase locked to the toggle field frequency, so that the output is proportional to $|M_z|$. Fig. 1(d) shows data for the virgin device, together with a fit to the model above yielding $H_k \approx 6000$ Oe.

The magnetic anisotropy was then modified by applying a gate voltage of -5 V to the Au gate electrode (Fig. 1(b)) for 10 minutes at a substrate temperature of 120°C. Elevated temperature is used to enhance the $O^{2-}$ vacancy mobility, so that under negative bias, oxygen vacancies drift toward the Au electrode and $O^{2-}$ is displaced towards the Co layer.[29] After removing the voltage and returning to room temperature, the device retained an out-of-plane easy axis with full remanence at zero field, but exhibited a sheared hysteresis loop (Fig. 1(c)), consistent with diminished PMA that is expected in the case of an over-oxidized Co interface.[29] Devices on the same wafer that experienced no voltage application showed no change in magnetic properties, indicating that the elevated temperature does not cause significant annealing effects.

$H_k$ was measured in the magneto-ionically modified state by measuring $M_z$ versus $H_x$ as described above (Fig. 1(d)). A reasonable fit to the Stoner-Wohlfarth model could not be attained, likely because of the onset of a multi-domain state at low $H_x$ so that $<M_z>$ drops more rapidly with $H_x$ than the model can account for. Here we estimate $H_k$ to be 1200 Oe, where the measured $M_z$ vanishes. The reflectivity of the device was also reduced by a factor of ~3, consistent with electrochromic discoloration due to ionic motion, suggesting that after voltage application the Co layer has become partially oxidized.

Polar MOKE was used to quantify the SOT effective fields before and after voltage application, through harmonic detection of low-frequency current-driven oscillation of $M_z$ analogous to the electrical approach in Refs. 10,26,30–33 using the anomalous Hall effect. Since the measurement timescale is much slower than precessional dynamics, the SOTs can be considered in terms of effective magnetic fields,[34,35] $\vec{H}_{SL} \sim \hat{m} \times (\hat{z} \times \hat{j}_e)$ for the Slonczewski-like SOT and $\vec{H}_{FL} \sim \hat{z} \times \hat{j}_e$ for the field-like SOT, where $\hat{m}$, $\hat{z}$, and $\hat{j}_e$ are the unit vectors in the directions of the magnetization, out-of-plane z-axis, and in-plane electron flow along the *x*-axis (Fig. 2), respectively. An in-plane AC current is injected through the thin-film microstrip, and the magnetization vector oscillates about its equilibrium direction due to the oscillating effective fields $H_{SL}$ and $H_{FL}$. A static in-plane field restricts the axis of this magnetization oscillation. A longitudinal field $H_x$ aligns the oscillation along the current axis, such that the detected oscillation in $M_z$ is due to $H_{SL}$ (Fig. 2(a)). A transverse field $H_y$ ensures that the detected oscillation is from $H_{FL}$ (Fig. 2(b)).

Magneto-optical detection has recently been used in Ref. 36 to probe field-like SOTs and by Fan *et al.* to quantify vector SOTs in in-plane magnetized bilayer thin films.[37] This optical detection method does not suffer from a thermoelectric (anomalous Nernst effect) artifact that may be substantial in the anomalous Hall voltage technique, nor does it suffer from signal contamination due to the planar Hall effect contribution.[32] In our experiments we use a normal incidence laser probe, so that the longitudinal and transverse MOKE signals (due to the in-plane magnetization components) are negligible and the detected signal thus serves as a sensitive probe of $M_z$.

An AC current was injected along the strip at angular frequency $\omega$ = 3142 rad/s, and the first harmonic $A_\omega$ of the polar MOKE signal was measured using a lock-in amplifier (Fig. 2(c)). Following a derivation similar to Ref. 31 (see Supplementary Material[38]) we obtain

$$\frac{A_\omega}{A_o} = \mp \frac{H_{SL(FL)}}{H_k^2} H_{x(y)}, \qquad (\text{Eq.1})$$

where $A_o$ is the maximum amplitude of the MOKE signal, i.e., half the difference between the MOKE signal levels for the up and down magnetized states. The "-" and "+" signs in front correspond to the up and down magnetized states, respectively.

Fig. 3 shows examples of the normalized harmonic MOKE signal $A_\omega/A_o$ as a function of quasi-statically swept in-plane fields $H_x$ and $H_y$, before and after magneto-ionic modification, at AC current density amplitude $|J_e|$ = 8.5×10$^{10}$ A/m$^2$. Note that the maximum in-plane field for the modified device was limited to 100 Oe to prevent domain nucleation and onset of a non-uniform magnetization state. While the polarities of the slopes remained the same after the magneto-ionic modification, the magnitudes of the slopes changed markedly. According to Eq. 1, this is due to the ~5-fold reduction in $H_k$ as noted above, as well as the changes in the SOT effective fields $H_{SL}$ and $H_{FL}$ quantified below.

Fig. 4 shows $H_{SL}$ and $H_{FL}$ versus $|J_e|$, computed via Eq. 1 with $H_k$ = 6000 Oe for the virgin device and $H_k$=1200 Oe for the modified device. The data are linear in $|J_e|$ within the measured current range. From the slopes we find $|H_{SL}|$ = 4.2 Oe and $|H_{FL}|$ = 1.2 Oe per 10$^{10}$ A/m$^2$ for the virgin state, and $|H_{SL}|$ = 46.8 Oe per 10$^{10}$ A/m$^2$, and $|H_{FL}|$ = 1.6 Oe per 10$^{10}$ A/m$^2$ in the modified state. The results for the virgin device are comparable to those reported elsewhere for similar Pt-based out-of-plane magnetized ultrathin ferromagnets.[10,32] Assuming that the Slonczewski-like torque arises entirely from the spin Hall effect, we can estimate the effective spin Hall angle $\theta_{SH}$ from[35,39]

$$\left|\frac{H_{SL}}{J_e}\right| = \frac{\hbar|\theta_{SH}|}{2|e|M_s t_F}, \qquad (\text{Eq. 2})$$

where $M_s \approx 1200$ emu/cm$^3$ is the saturation magnetization and $t_F = 0.9$ nm is the nominal ferromagnet thickness. The resulting $\theta_{SH} \approx 0.14$ for Pt in the virgin Pt/Co/GdOx device is comparable to the values obtained with other techniques.[3,10,12,32,37,40–42]

After the magneto-ionic modification, the estimated Slonczewski-like effective field increases by approximately an order of magnitude compared to the virgin-state device. Based on this increase in $|H_{SL}|$ and the fivefold decrease in $H_k$, an increase in SOT switching efficiency by a factor of ~50 would be anticipated. From Eq. 2, a reduction of $M_s t_F$ due to partial Co oxidation could potentially account for the large voltage-induced enhancement of $|H_{SL}|$. Taking the reduction of the polar MOKE signal as a gauge of oxidation, we find in Fig. 1(c) that the ratio of the polar MOKE signal intensity $\Delta I$ to the mean reflected intensity $I_0$ decreases from $\Delta I/I_0 = 1.74 \times 10^{-2}$ in the virgin state to $\Delta I/I_0 = 0.89 \times 10^{-2}$ in the magneto-ionically modified state. Assuming the magneto-optical properties remain unchanged by voltage, this reduction in the normalized MOKE signal by a factor of 2 implies a corresponding factor of 2 decrease in $M_s t_F$, which alone does not fully account for the increase in $|H_{SL}|$. We note that the model used to derive Eq. 1 (Refs. 31 and 38) assumes the magnetization of the sample to be uniform. This assumption may not be valid in the case of the magneto-ionically modified device with reduced PMA, and inhomogeneity in the magnetization may result in a large apparent $|H_{SL}|$ extracted from the harmonic technique.[10]

These possible origins for the significantly increased value for $|H_{SL}|$ might be expected to similarly enhance the value of $|H_{FL}|$. However, the measured field-like torque remains at a similar magnitude even after the magneto-ionic modification, such that the estimated ratio

$|H_{SL}/H_{FL}|$ increases by as much as an order of magnitude from ≈3 to ≈30. This remarkable change in the relative SOT magnitudes could result from a change in the electronic potential at the Co/GdOx interface due to the change in Co-oxygen orbital hybridization, which would be expected to influence any Rashba-like contributions to SOTs at that interface.[17] Given the small layer thicknesses in this system, a change in interface structure might also influence charge and spin transport by modifying interfacial scattering rates, which would be expected to affect the current-induced SOTs.[34]

In summary, we have demonstrated non-volatile control over SOTs through voltage-induced oxygen ion migration at the Co/GdOx interface. Together with the previously described strong change in PMA,[29] we observe a significant enhancement of the Slonczewski-like torque. Since it has been shown elsewhere that the interfacial oxidation state can be reversibly controlled with a gate voltage in this manner,[28,29] these results imply that a gate voltage can be used to toggle a device between a "writing" state for low-energy SOT switching and a "retention" state with higher anisotropy for long-term stability. The observed increase in the ratio $|H_{SL}/H_{FL}|$ suggests that changes to the oxidation state at the Co/oxide interface leads to nontrivial changes in the SOTs. Although the origin of the modified SOTs remains to be explored in detail, the mechanism we present here to tune the interface provides a route to increase the efficiency of electrically-driven magnetization switching while gaining insight into the microscopic processes responsible for SOTs.

This work was supported in part by C-SPIN, one of the six SRC STARnet Centers, sponsored by MARCO and DARPA, and by the National Science Foundation under NSF-ECCS-1128439. Work was performed using instruments in the MIT Nanostructures Laboratory, the Scanning

Electron-Beam Lithography facility at the Research Laboratory of Electronics, and the Center for Materials Science and Engineering at MIT.

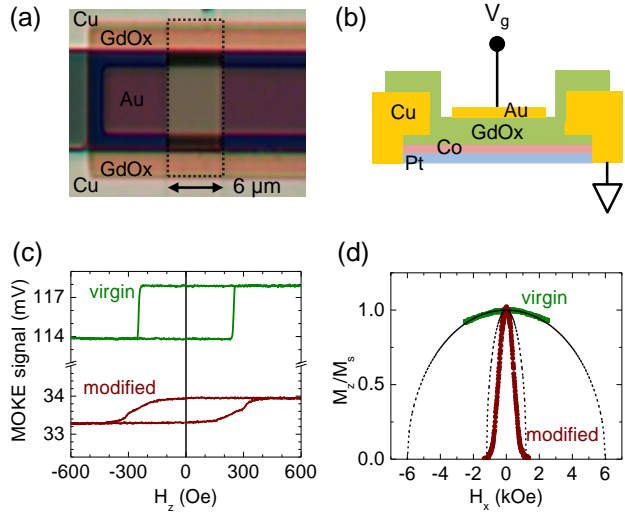

figure 1. (a) Optical micrograph of the gated Pt/Co/GdOx device. The Pt/Co/GdOx microstrip is enclosed by the dotted outline. (b) Schematic of the cross-section of the device in (a). (c) Out-of-plane hysteresis loops, obtained with the magneto-optical Kerr effect (MOKE), of the Pt/Co/GdOx device in the virgin state and after magneto-ionic modification. (d) Normalized out-of-plane component of the magnetization $M_z/M_s$ measured against in-plane field with MOKE for the device in virgin state and after magneto-ionic modification. The dotted curves are fits to the single-domain model for estimating the in-plane saturation field $H_k$.

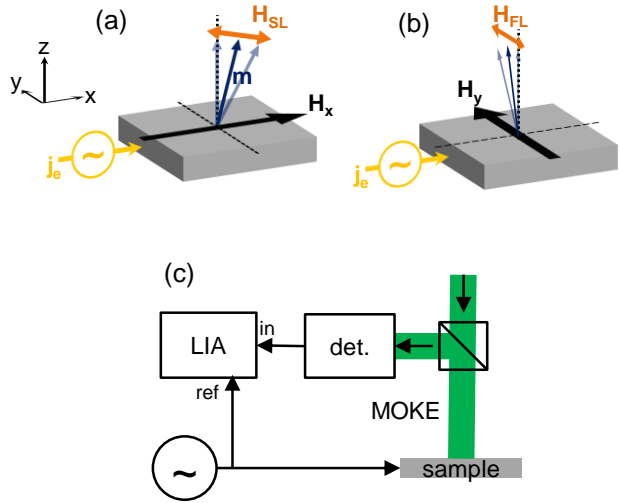

Figure 2. Illustrations of effective fields associated with the (a) Slonczewski-like (damping-like) torque $H_{SL}$ and (b) field-like torque $H_{FL}$. (c) Schematic of spin-orbit torque detection using the magneto-optical Kerr effect using a lock-in amplifier (LIA).

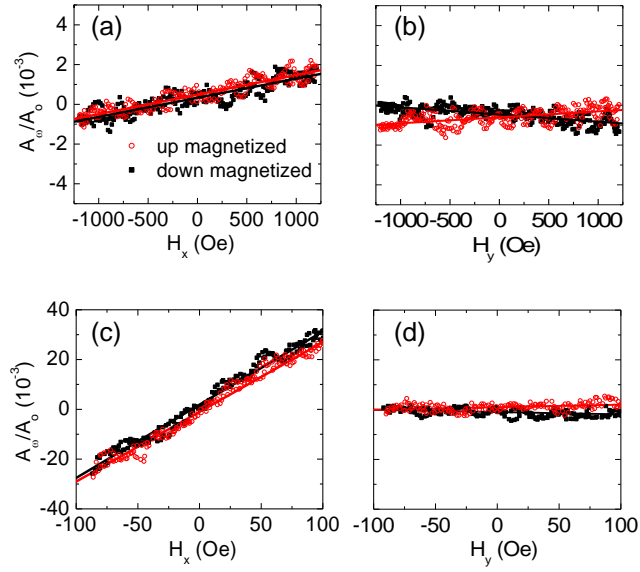

Figure 3. Harmonic MOKE signal versus in-plane fields $H_x$ and $H_y$ for Pt/Co/GdOx (a, b) in the virgin state and (c, d) after magneto-ionic modification. The amplitude of the AC electron current density $|J_e|$ is $8.5 \times 10^{10}$ A/m$^2$.

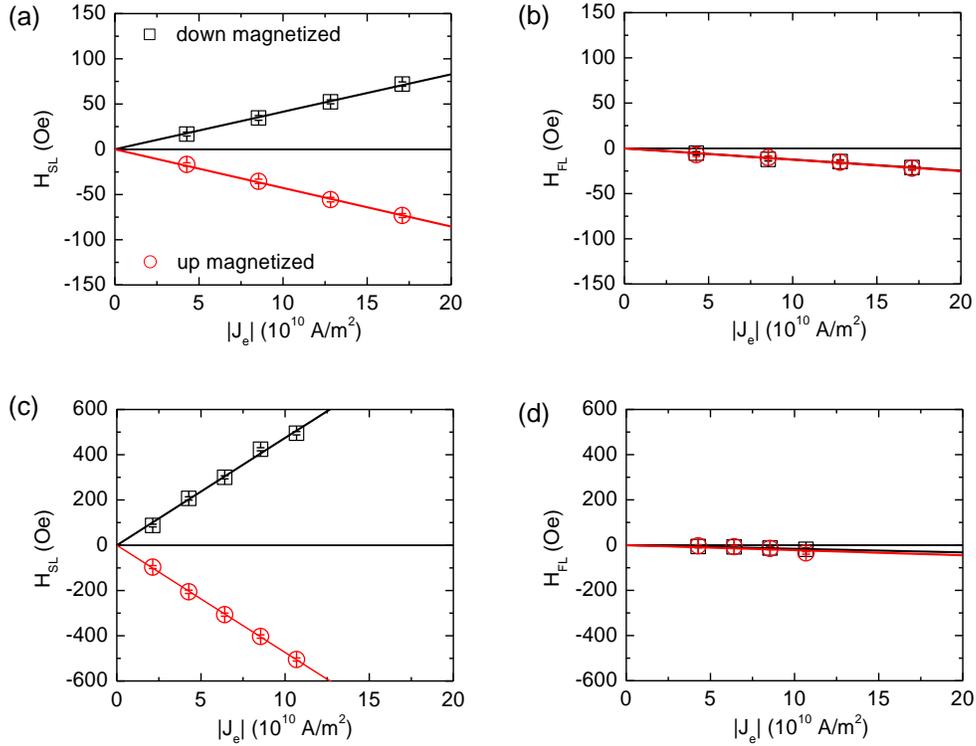

Figure 4. Effective fields $H_{SL}$ and $H_{FL}$ due to Slonczewski-like and field-like spin-orbit torques, respectively, against electron current density amplitude $|J_e|$ for Pt/Co/GdOx (a, b) in the virgin state and (c, d) after magneto-ionic modification.